\documentclass[aps,twocolumn,prb,reprint,amsmath,amssymb,showpacs,superscriptaddress]{revtex4-1}
\usepackage{multirow}
\usepackage{graphicx,epsfig}
\usepackage[colorlinks=true,citecolor=blue,linkcolor=red,linktocpage=true,pagebackref=false]{hyperref}
\usepackage[linktocpage=true]{hyperref}

\newcommand{\ba}{\begin{eqnarray}}
\newcommand{\ea}{  \end{eqnarray}}
\newcommand{\ve}{\varepsilon}

\def \beq {\begin{equation}}
\def \edq {\end{equation}}
\def \bes {\begin{subequations}}
\def \eds {\end{subequations}}
\def \beqn {\begin{equation*}}
\def \edqn {\end{equation*}}

\begin{document}
\title{Adiabatic response and quantum thermoelectrics for {\em ac} driven quantum systems}
\author{Mar\'{\i}a Florencia Ludovico}
\affiliation{Departamento de F\'{\i}sica, FCEyN, Universidad de Buenos Aires and IFIBA, Pabell\'on I, Ciudad Universitaria, 1428 CABA Argentina}
\affiliation{International Center for Advanced Studies, UNSAM, Campus Miguelete, 25 de Mayo y Francia, 1650 Buenos Aires, Argentina}
\author{Francesca Battista}
\affiliation{Departamento de F\'{\i}sica, FCEyN, Universidad de Buenos Aires and IFIBA, Pabell\'on I, Ciudad Universitaria, 1428 CABA Argentina}
\affiliation{International Center for Advanced Studies, UNSAM, Campus Miguelete, 25 de Mayo y Francia, 1650 Buenos Aires, Argentina}
\author{Felix von Oppen}
\affiliation{\mbox{Dahlem Center for Complex Quantum Systems and Fachbereich Physik, Freie Universit\"at Berlin, 14195 Berlin, Germany}}
\author{Liliana Arrachea}
\affiliation{Departamento de F\'{\i}sica, FCEyN, Universidad de Buenos Aires and IFIBA, Pabell\'on I, Ciudad Universitaria, 1428 CABA Argentina}
\affiliation{International Center for Advanced Studies, UNSAM, Campus Miguelete, 25 de Mayo y Francia, 1650 Buenos Aires, Argentina}

\begin{abstract}
We generalize the theory of thermoelectrics to include coherent electron systems under adiabatic {\em ac} driving, accounting for quantum pumping of charge and heat as well as for the work exchanged between electron system and driving potentials. We derive the relevant response coefficients in the adiabatic regime and show that they obey generalized Onsager reciprocity relations. We analyze the consequences of our generalized thermoelectric framework for quantum motors, generators, heat engines, and heat pumps, characterizing them in terms of efficiencies and figures of merit.  We illustrate these concepts in a model for a quantum pump.
\end{abstract}

% 73.23.-b Electronic transport in mesoscopic systems 
% 72.10.Bg General formulation of transport theory 
% 73.63.Kv Quantum dots 
% 44.10.+i Heat conduction

\maketitle

\section{Introduction}
Describing the relation between particle and energy currents is at the heart of thermoelectrics. \cite{gen, revcasati,ben, seif, ora} For {\em dc} driving with small temperature gradients and bias voltages, linear-response relations between the currents and the applied forces constitute the basis to describe thermoelectric phenomena. When combined with the principles of thermodynamics, the resulting theory has the beauty of simplicity and the strength of high predictive power. Specifically, it allows for a successful characterization of the efficiency of various thermoelectric machines in terms of the figure of merit introduced by Ioffe in 1949.\cite{zt}

An important challenge is to incorporate genuine quantum effects associated with coherent transport in nano\-devices into this theoretical framework for thermoelectric effects. Here, we address how to include adiabatic quantum pumping as a paradigm of coherent-transport effects into a suitably generalized thermoelectric framework and explore the fundamental relations of the corresponding quantum machines. Quantum pumping generates nonzero {\em dc} currents by locally applying  purely {\em ac} drivings to a quantum coherent conductor \cite{pumps, BPT, brou}. It generates both charge and energy currents \cite{Moskalets04}, enables heat pumping, and the exchange of work between different driving forces \cite{arr07,mos09}. The aim of the present work is to extend the linear-response theory of thermoelectric effects to systems under adiabatic driving. To this end,  we need to include the energy flux between the electrons and the {\em ac} forces on an equal footing with the heat and particle fluxes. 

\begin{figure}[b]
  \includegraphics[scale=0.45]{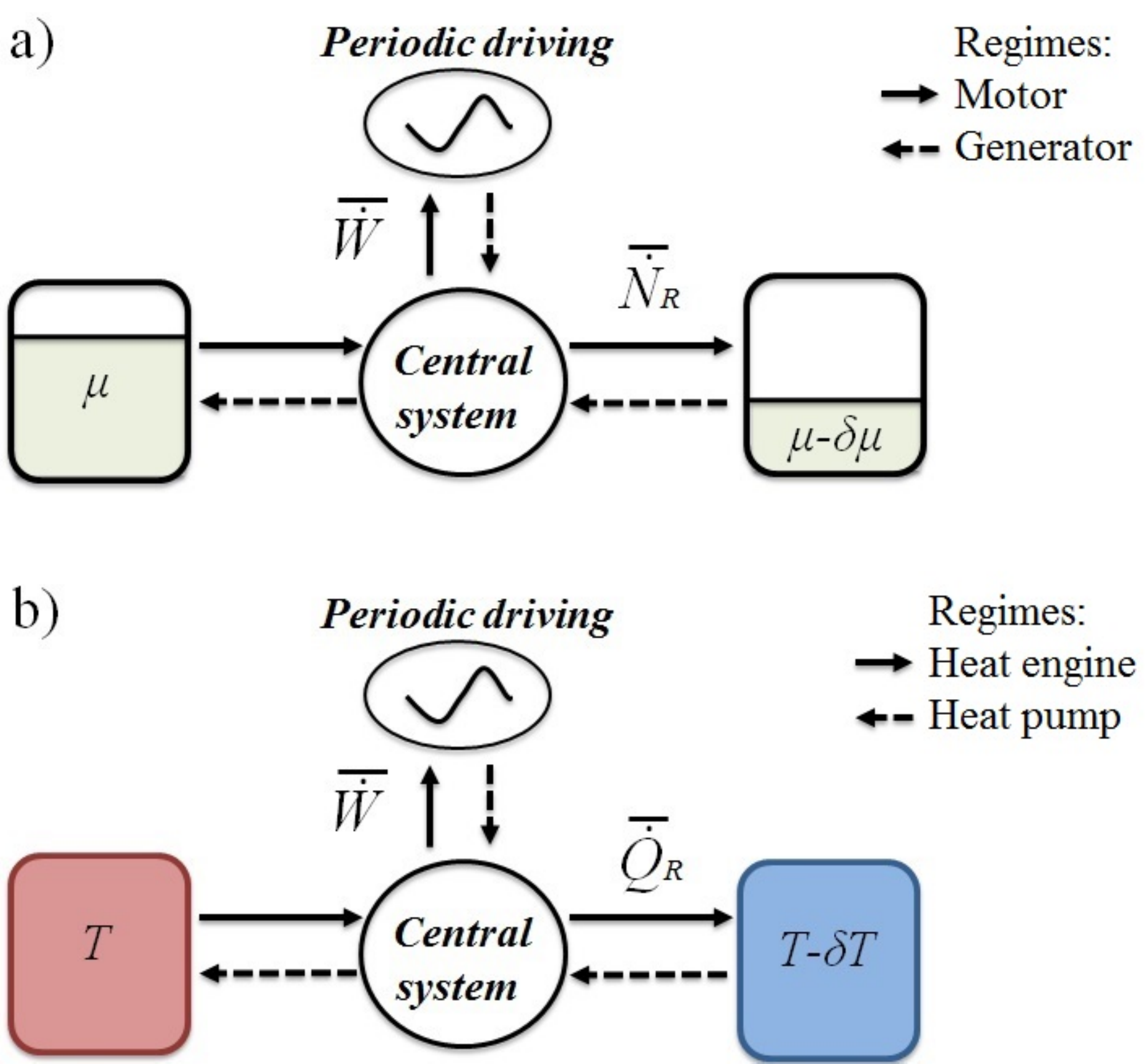}
  \caption{Sketch of the setup. A coherent quantum conductor is driven by time-periodic potentials and connected to two reservoirs biased by (a) a chemical-potential difference $\delta\mu$ or (b) a temperature gradient $\delta T$, or both. Charge $\overline{\dot{N}}_{R}$, heat $\overline{\dot{Q}}_{R}$ and power $\overline{\dot{W}}$ are exchanged between the reservoirs and the {\em ac} sources. The solid (dashed) arrow indicate (a) the motor (generator) mode of the device, and (b) the heat engine (heat pump) mode.}\label{setup}
\end{figure}

Figure \ref{setup} shows the setup that we have in mind. It consists of a central coherent conductor which is coupled to two reservoirs. In conventional thermoelectrics,\cite{revcasati} the two reservoirs differ in both  temperature and chemical potential. A thermal engine converts a temperature difference into electric power. As a consequence of the Second Law, the efficiency of this conversion process is limited by the Carnot efficiency $\eta_C = (T_2-T_1)/T_2$ where $T_2>T_1$ denote the temperatures of the reservoirs. The optimal efficiency that can be reached for a specific device is controlled by its figure of merit or ZT value,\cite{revcasati} 
\begin{equation}
  \eta = \eta_C \frac{\sqrt{1+ZT}-1}{\sqrt{1+ZT}+1}.
\label{maxeff}
\end{equation}
The ZT value can be expressed in terms of the linear-response coefficients of the device, relating charge and heat currents to bias voltage and temperature gradient.\cite{revcasati} 

The thermal engine can also be operated in reverse, realizing a refrigerator which invests electric power to continuously extract heat from the colder reservoir. The maximal efficiency of this device is given by the appropriate Carnot efficiency $\eta_C=T_1/(T_2-T_1)$ and in terms of this Carnot efficiency, the optimal efficiency for a specific device is again given by Eq.\ \ref{maxeff}.\cite{revcasati} 

In this paper, we consider setups in which the coherent conductor is subject to a set of {\em ac} potentials in addition. For definiteness, we will consider reservoirs which have either different chemical potentials [Fig.\ \ref{setup}(a)] or different temperatures [Fig.\ \ref{setup}(b)], although our theory could readily be applied to situations which combine {\em ac} potentials with both chemical-potential and temperature gradients. 
The physics of these setups can be understood by analogy to the Archimedes device, a pipe with a rotating screw, which can be used to pump water against gravity. This is a classical analog of an adiabatic quantum pump, where {\em ac} driving pumps a certain amount of electric charge per cycle. Specifically, this charge can be pumped against an applied {\em dc} bias voltage,\cite{Janine} in which case quantum pump realizes a generator. 

The Archimedes screw can also be operated in reverse, with water flowing between the reservoirs by gravity and setting the screw into rotational motion. An analogous effect can be used to turn an adiabatic  quantum pump into an adiabatic quantum motor. This is most easily understood when imagining that the time dependence of the {\em ac} potentials derives from the (classical) dynamics of, say, one or more mechanical degrees of freedom.\cite{bus} Then, a charge current pushed through the coherent conductor will set the mechanical degrees of freedom into motion. 

Generator and quantum motor are driven by a bias voltage and correspond to the setup sketched in Fig.\ \ref{setup}(a). Alternatively, we can also consider devices involving temperature gradients instead of bias voltages, which realize heat pumps and heat engines. Such a device is depicted in Fig.\ \ref{setup}(b). The devices in Fig.\ \ref{setup} are examples of nanomotors and nanoengines, which have received much attention recently.\cite{bus,nanomot,nanoeng, nanomachine, Linke02, Linke05} We note that the effect of {\em ac} potentials on the conventional thermoelectric effects has been studied in a number of recent papers.\cite{crepieux,Janine,dsanchez,dare} 

The aim of the present work is to extend the linear-response theory of thermoelectrics to such nanomotors and nanoengines, to understand their efficiencies, and to identify appropriate figures of merit. This program poses several conceptual questions: (i) We need to identify the current that complements the charge and heat currents and accounts for the effects of the {\em ac} potentials. Similarly, we need to identify the affinity that complements the (scaled) temperature difference and bias voltage. (ii) We need to develop the generalized linear response theory which includes these additional quantities. While this is a conventional linear-response theory for traditional thermoelectrics, the {\em ac} potentials are not actually weak but only slowly varying. (iii) We finally need to identify appropriate efficiencies and figures of merit. We will see that the latter also differ in essential ways from those defined in conventional thermoelectrics. 

In Sec.\ \ref{secadia}, we generalize linear-response theory to include the response to the adiabatically varying {\em ac} potentials in addition to the applied bias voltage. We do this by working to linear order in the rate of change (or velocity) of the {\em ac} potentials. We find that this can be done in a manner which closely resembles the derivation of Kubo formulas in linear response theory. Consequently, we derive general Kubo-like expressions for the response of both the charge current and the generalized forces conjugate to the {\em ac} potentials. These expressions imply that the response coefficients satisfy Onsager-like relations and are thus not independent of one another. In Sec.\ \ref{secframe},  we generalize the thermodynamical framework to include the time-averaged work per unit time performed by the {\em ac} forces as a third flux, along with the heat and particle fluxes. We also identify the scaled frequency $\hbar\omega/T$ of the driving as the appropriate third affinity, complementing the temperature and chemical-potential differences. In Sec.\ \ref{seceff}, we  define and analyze efficiency and figure of merit for the various quantum machines sketched in Fig.\ \ref{setup}. We find that the definition of the appropriate figure of merit analogous to the ZT value differs in characteristic ways, reflecting the fact that the usual off-diagonal thermoelectric response coefficients, the off-diagonal coefficients involving the third flux or affinity do not enter into the entropy production. To illustrate these concepts, we apply our theory to an example device in Sec.\ \ref{secex}  which is based on a simple model for a quantum pump. We summarize in Sec.\ \ref{seccon}.
 
\section{Adiabatic response and Onsager relations}
\label{secadia}

We begin by evaluating the forces and currents induced by a set of time-periodic parameters in the adiabatic approximation. We will see that this can be done in close analogy to linear-response theory, allowing us to derive Onsager-like relations.

We collect the parameters $V_i(t)$ of the Hamiltonian $\hat{{\cal H}}$ into a vector ${\bf V}(t)= {\bf V}(t+{\cal T}) = \left( V_1(t), V_2(t), \ldots  \right)$ so that $\hat{{\cal H}}=\hat{{\cal H}}({\bf V}(t))$, where ${\cal T}=2
\pi/\omega$ is the driving period.\cite{footnote} Quite generally the Hamiltonian of the system can be expressed as 
\begin{equation}
 \hat{{\cal H}}({\bf V(t)})=  \hat{{\cal H}}_0 - \sum_{j} \hat{F}_j V_j(t),
 \end{equation}
where ${\cal H}_0$ is the time-independent part of the Hamiltonian and ${F}_j$ are hermitian operators that play the role of generalized forces
\begin{equation}
    {\bf \hat{F}}(t)=-\frac{\partial \hat{{\cal H}}(t)}{\partial {\bf V}(t)}.
\end{equation}
The quantum expectation value ${\rm tr} \{ \hat \rho {\bf\hat F}\} $ in terms of the electronic density matrix $\hat \rho$ are just conventional forces when the $V_j$ denote regular cartesian coordinates of a classical system obeying Newtonian dynamics. 

At lowest order in the adiabatic approximation, the system is described by the frozen density matrix $\hat{\rho}_t $ for the Hamiltonian $\hat{{\cal H}}_t$ with $t$ treated as a parameter. Accounting for the temporal variation of ${\bf V}(t)$ to lowest order, we can approximate the time evolution operator as 
\begin{equation}
\hat{U}(t,t_0) \simeq {\mathrm T} \exp\{-i \hat{{\cal H}}_t (t-t_0) - i \int_{t_0}^t dt^{\prime} (t-t^{\prime}) {\bf \hat{F}} \cdot \dot{\bf V}(t) \}.
\end{equation}
 To linear order in the small ''velocity'' $\dot{\bf V}(t)$, we can now follow the usual steps of linear response theory \cite{bruus} and express the expectation value $O(t)$ of an observable $\hat{O}$ at time $t$ as
\begin{eqnarray}
O(t)&\simeq& \langle \hat{O} \rangle_t  - i \int_{t_0}^t dt^{\prime} (t-t^{\prime}) \langle \left[\hat{O}(t), \hat{\bf F}(t^{\prime})     
     \right]\rangle_t \dot{\bf V}(t) \nonumber\\
     &=& \langle \hat{O} \rangle_t  + {\bf \Lambda}^{O {\bf F}}_t \cdot \dot{\bf V}(t).
\end{eqnarray}
Here, the operators $\hat{O}(t)$ and $\hat{\bf F}(t^{\prime})$ are defined in the Heisenberg representation with respect to the frozen Hamiltonian  ${\cal H}_t$ and 
$\langle\ldots \rangle_t$ denotes the expectation value with respect to the frozen density matrix $\hat{\rho}_t$. The response function ${\bf \Lambda}^{O {\bf F}}_t $ can be expressed through the retarded adiabatic  susceptibility $ \chi^{O,{\bf F}}_t (t-t^{\prime})= -i \theta(t-t^{\prime}) \langle [\hat{O}(t), \hat{\bf F}(t^{\prime})]\rangle_t$. We now expand the frozen average to linear order in an applied bias $\delta\mu$, yielding $\langle\hat{O} \rangle_t \simeq  \Lambda^{O c}_t \delta\mu$, where the linear-response coefficient $\Lambda^{O c}_t$ is given by the usual  Kubo formula. Applying this procedure specifically to the charge current $J^c(t)$ and the forces ${\bf F}(t)$  (and postponing the heat currents and temperature gradients for further below), we obtain   
\begin{equation}\label{linear}
    \left( \begin{array}{c}  J^c(t) \\ {\bf F}(t) \end{array} \right) =  \left( \begin{array}{c} 
     J^c_t  \\ {\bf F}_t  \end{array} \right) + \left( \begin{array}{cc} \Lambda^{cc}_t & {\bf \Lambda}^{cf}_t \\
     {\bf \Lambda}^{fc}_t &   \hat{\bf \Lambda}^{ff}_t\end{array} \right)  \left( \begin{array}{c} \delta \mu \\
   \dot{\bf V}(t) \end{array} \right),  
\end{equation}
to linear order in $\delta \mu$ and $\dot{\bf V}(t)$. 

The terms in Eq.\ (\ref{linear}) have clear physical interpretations. The first term on the right hand side collects the currents and forces evaluated with the frozen density matrix $\hat{\rho}_t$ in equilibrium (\textit{i.e.}, for $\delta\mu=0$). These terms have zero mean when averaged over one period of the {\em ac} fields. The forces can be thought of as conservative Born-Oppenheimer forces and expressed as a gradient of the equilibrium energy of the system with respect to ${\bf V}(t)$. For several potentials this term may lead to exchange of work between the different forces $F_j$ without dissipation. Such processes were considered in Refs.\ \onlinecite{arr07} and \onlinecite{mos09}. Adiabatic quantum pumping of charge by the {\em ac} potentials is described by ${\bf \Lambda}^{cf}_t$, while ${\bf \Lambda}^{fc}_t$ captures the modification of the forces by the applied bias $\delta\mu$.  Both contributions are generally nonzero when averaged over a period, implying that this contribution to the force is {\em non}conservative. This was discussed  for non-interacting electrons coupled to adiabatic nanomechanical systems  \cite{brand,niels1} and nanomagnets. \cite{niels2} In the latter case, this corresponds to a spin-transfer torque. The diagonal components describe the usual conductivity through $\Lambda^{cc}_t$ and the velocity-dependent force through $\hat{\bf \Lambda}^{ff}_t$. In time-reversal symmetric systems, the latter is symmetric and describes a frictional force. Without time-reversal symmetry, $\hat{\bf \Lambda}^{ff}_t$ may have an antisymmetric part which is analogous to the Lorenz force. \cite{niels1}

The derivation of the response coefficients $\Lambda_t^{ij}$ follows standard linear-reponse theory, including the ``adiabatic" response to the {\em ac} potentials. Consequently, it is natural to expect that the response functions $\Lambda_t^{ij}$ satisfy Onsager-like relations. In fact, these can be derived in the usual manner, as shown in detail in App.\ \ref{apon}. Thus, we find the generalized Onsager relations
\begin{eqnarray} \label{onslam}\nonumber
\Lambda_t^{cc}(B)= \Lambda_t^{cc}(-B) \,\,&,&\,\, {\Lambda}_{ij}^{ff}(B)= s_i s_j { \Lambda}_{ji}^{ff}(-B)\\ 
{\Lambda}_j^{cf}(B) &=& s_j { \Lambda}_j^{fc}(-B),
\end{eqnarray}
where the sign $s_j=\pm$ depends on the parity of the operators $\hat{F}_j$ under time reversal. As the derivation of Onsager relations is very general, these relations are valid at finite temperature $T$ and in the presence of many-body interactions.\cite{cohen} 

The second line in Eq.\ (\ref{onslam}) imposes a relation between the adiabatic quantum pumping of charge (as described by ${\Lambda}_j^{cf}$) and the nonconservative force (as described by ${\Lambda}_j^{fc}$). This relation which is valid in the adiabatic regime was previously found for noninteracting adiabatic quantum motors at zero temperature and $B=0$.\cite{bus} It has been pointed out that time-reversal symmetry -- by way of Onsager-like arguments -- does not imply symmetry of the pumped charge under magnetic-field reversal unless the system has additional spatial symmetries.\cite{sim1,sim2,sim3,sim4,sim5,ludo2} The relation in Eq.\ (\ref{onslam}) implies that there is still an Onsager relation associated with the pumped charge, but it does not relate the pumped charge to itself but rather to the nonconservative force in response to an applied bias.

Before closing this section, we comment on how to include heat currents and thermal gradients into this linear-response scheme. Within linear response, we can readily extend Eq.\ (\ref{linear}) into a $3\times3$ matrix equation 
\begin{equation}\label{linear2}
    \left( \begin{array}{c}  J^c(t) \\ J^Q (t) \\ {\bf F}(t) \end{array} \right) =  \left( \begin{array}{c} 
     J^c_t \\ J^Q_t \\ {\bf F}_t  \end{array} \right) + \left( \begin{array}{ccc}   \Lambda^{cc}_t &  \Lambda^{cq}_t & {\bf \Lambda}^{cf}_t \\
      \Lambda^{qc}_t &  \Lambda^{qq}_t & {\bf \Lambda}^{qf}_t \\
     {\bf \Lambda}^{fc}_t &  {\bf \Lambda}^{fq}_t &   \hat{\bf \Lambda}^{ff}_t\end{array} \right)  \left( \begin{array}{c} \delta \mu \\ \delta T \\
   \dot{\bf V}(t) \end{array} \right).
\end{equation}
Here, we can identify the thermal conductance $ \Lambda^{qq}_t $ relating the heat current $J^Q(t)$ to $\delta T$ as well as the usual thermoelectric coefficients $\Lambda^{cq}$ and $\Lambda^{qc}$. In addition, our scheme includes the coefficients ${\bf \Lambda}^{qf}_t$ and ${\bf \Lambda}^{fq}_t$ which describe the generation of heat currents by a time-dependent driving (quantum pumping of heat) and the generation of a nonconservative force in response to a temperature gradient, respectively.

The treatment of a temperature gradient within the Kubo approach is less straightforward, but has been addressed numerous times in the literature.\cite{mahan,lutt,shastry} An alternative route is to calculate the relevant observables with a non-equilibrium technique, such as the Keldysh or scattering matrix formalisms, and to perform the expansions in $\delta T, \delta \mu$, and $\dot{\bf V}$ {\em a posteriori}. (This is the route which we follow in Sec.\ V). Either approach yields the additional Onsager relations
\begin{eqnarray} \label{onslamad}\nonumber
\Lambda_t^{qq}(B)= \Lambda_t^{qq}(-B) \,\,&,&\,\, {\Lambda}_t^{cq}(B)= { \Lambda}_{t}^{qc}(-B) \\ 
{\Lambda}_j^{qf}(B) &=& s_j  \Lambda_j^{fq}(-B)
\end{eqnarray}
complementing Eq.\ (\ref{onslam}).
The first line corresponds to the usual thermoelectric Onsager relations. The second line contains the additional Onsager relations relating pumping of heat current and the force generated in response to an applied thermal gradient.

\section{Generalized thermoelectric framework}\label{secframe}

Conventional thermoelectrics considers particle and heat currents in response to chemical-potential and temperature differences. In the presence of {\em ac} driving as in the devices in Fig.\ \ref{setup}, we have to take into account the pumping of particles and heat as well as the work performed by or on the {\em ac} potentials on the same footing. To develop the corresponding generalized thermoelectrics, we first consider the entropy production of the system. After averaging over one period of the {\em ac} driving, the net dissipation occurs only in the electrodes and we can write
\begin{equation} 
 \overline{\dot{S}}= \overline{\frac{{\dot{Q}}_L}{T_L}}+\overline{\frac{{\dot{Q}}_R}{T_R}},
\label{sac}
\end{equation}
where the average heat flux in lead $\alpha$ is given by 
\begin{equation}
  \overline{\dot{Q}}_{\alpha} = {\overline{\dot{E}}_{\alpha}} - \mu_{\alpha} {\overline{\dot{N}}_{\alpha}}.
\label{heat} 
\end{equation}
The energies $E_\alpha$ and particle numbers $N_\alpha$ satisfy the conservation laws
\begin{eqnarray} \label{cons}
& & \overline{\dot{N}}_R  =   - \overline{ \dot{N}}_L , \;\;\;\;\;\  \overline{\dot{E}}_L + \overline{\dot{E}}_R = \overline{\dot{W}}.
\end{eqnarray}
While particle-number conservation takes the same form as in standard thermoelectrics, energy conservation must account for the additional work $W$ performed by the {\em ac} potentials on the electron system. The corresponding power can be expressed as ${\dot{W}} = -\sum_j {F_j(t) \dot{V}_j(t)}$, yielding the entropy production
\begin{equation}
 \overline{\dot{S}}\!=\!\overline{\dot N}_R \frac{\delta\mu}{T}\!+\!\overline{\dot Q}_R\frac{\delta T}{T^2}\!-\!\sum_j \overline{F_j(t)\frac{{\dot V}_j(t)}{T}}
\label{Sdot}
\end{equation}
to linear order in the applied bias $\delta\mu=\mu_L-\mu_R$ and temperature difference $\delta T=T_L-T_R$. Note that after averaging over a period, the conservative Born-Oppenheimer forces in Eq.\ (\ref{linear}) do not contribute to entropy production. Then, the power can be expressed in linear response and for $\delta T=0$  as 
\begin{eqnarray} 
 \label{wdot}
  \overline{\dot{W}}\!=\!-\!\sum_{j}\!\!\left(\overline{(\hat{\bf \Lambda}_t^{fc})_{j}\dot{V}_j(t) }\delta\mu\!+\!\sum_{l}\overline{(\hat{\bf \Lambda}_t^{ff})_{jl}\dot{V}_j(t)\dot{V}_l(t)}\right)\!.
\end{eqnarray}
Here, the first term on the right-hand side describes the work performed by the nonconservative force originating from the applied voltage $\delta\mu$ ($\delta T$ would contribute a similar term) and the second term is the dissipated power due to a frictional force on the {\em ac} potentials. 

In conventional thermoelectrics, one defines the particle and heat fluxes $J_1=\overline{ \dot{N}}_R$ and $J_2= \overline{\dot{Q}}_R$ as well as the corresponding affinities ${X}_1=\delta \mu/T$ and ${X}_2=\delta T/T^2$. \cite{callen,landau}  To extend thermoelectrics to the present situation, we need to identify appropriate fluxes and affinities for the {\em ac} driving terms. 

At first sight, Eq.\ (\ref{Sdot}) may suggest to define the $-F_j$ as fluxes and the ${\dot V}_j/T$ as the associated affinities. However, Eq.\ (\ref{Sdot}) holds only after averaging over one period. Before time averaging, the conservation laws involve additional terms \cite{ludo1} and the forces $F_j(t)$ contain contributions that are conservative. We can identify an appropriate affinity by noting that after averaging, the first term in Eq.\ (\ref{wdot}) is proportional to $\omega$, while the second term is proportional to $\omega^2$. It is thus natural to define the affinity $X_3= \hbar \omega/T$ with associated flux $J_3 = {\overline {\dot W}}/(\hbar \omega)$. \cite{callen,landau} Thus, Eq.\ (\ref{Sdot}) yields
\begin{equation} 
\label{dots}
   \overline{\dot{S}}=  \sum_j {J}_j {X}_j
\end{equation}
for the rate of entropy production. 

We complete our quantum thermoelectrics scheme by linear-response relations between fluxes and affinities, 
\begin{equation}
\label{ont}
   J_i = \sum_k L_{ik} X_k.
\end{equation}
The linear-response coefficients $L_{ij}$ are readily related to the coefficients which appeared in Eq.\ (\ref{linear2}). Indeed, we have
\begin{eqnarray}
L_{11}&=& T \overline{\Lambda_t^{cc}}, \;\;\;L_{12} = T ^2\overline{\Lambda_t^{cq}},\;\;\;L_{13}  =  T\overline{ {\bf \Lambda}_t^{cf} \cdot {\bf v}}, \nonumber \\
L_{21} & = &T \overline{\Lambda_t^{qc}}, \;\;\;L_{22} = T^2 \overline{\Lambda_t^{qq}},\;\;\;L_{23}  =  T \overline{ {\bf \Lambda}_t^{qf} \cdot {\bf v}}, \\
L_{31} & = & - T \overline{ {\bf \Lambda}_t^{fc} \cdot {\bf v}},\;\; L_{32}  =  - T^2 \overline{ {\bf \Lambda}_t^{fq} \cdot {\bf v}}, \;\;
\nonumber\\
&&\,\,\,\,\,\,\,\,\,\,\,\,\,\,\,\,\,\,\,\,\,\,\,\,\,\,\,\,\,\,\,\,\,\,\,\,\,\,
L_{33}  =  - T \overline{ {\bf v}^T \cdot \hat {\bf \Lambda}_t^{ff} \cdot {\bf v}}, \nonumber 
\end{eqnarray}
where we defined ${\bf v}$ through $\dot{\bf V}= \hbar \omega {\bf v}$ and  ${\bf v}^T $ denotes the transpose of ${\bf v}$. 

Thus, the coefficients $L_{ij}$ also obey Onsager relations, namely
\begin{equation}\label{BL}
      L_{ii}(B)=L_{ii}(-B) \;\; , \;\; L_{ij}(B)=\pm L_{ji}(-B), \;
\end{equation}
with $i\neq j$. The sign in the second relation depends on the behavior of the fluxes under time reversal. Assuming time reversal from now on (and thus $B=0$), this yields the relation $L_{12}=L_{21}$, which is well known from the usual theory of thermoelectrics, as well as $L_{13}=-L_{31}$ and $L_{23}=-L_{32}$. It is important to note that the offdiagonal response coefficients have the same sign in conventional thermoelectrics, while they have opposite signs when either $J_3$ or $X_3$ is involved. We will see that below this has significant consequences for the definition of figures of merit for the devices in Fig.\ \ref{setup}.
 
The transport coefficients $L_{ij}$  can be directly calculated from the coefficients $\Lambda$, which are in turn given in terms of the susceptibilities $\chi_t(\omega)$. Another possibility is to start from the expressions for the charge, heat, and work currents, to perform the expansions in $\hbar\omega$, $\delta \mu$, and $\delta T$, and to identify the coefficients $L$ from the resulting expressions. For noninteracting systems, this procedure is rather straightforward. 

In Sec.\ \ref{secex}, we will illustrate our general theory for a general noninteracting model of a two-terminal conductor and evaluate the various response coefficients explicitly. This will rely on Green function \cite{arr07,arr06,ludo2} and scattering matrix \cite{arr07,Moskalets04} expressions for the response coefficients which we derive by the procedure described in the previous paragraph. Details of the calculations are given in App.\ \ref{applinres}. The calculations start with the expressions for charge current [Eq.\ (\ref{charge})], heat current [Eq.\ (\ref{energy})], and work current [Eq.\ (\ref{work})] for this model. Performing the expansions in $\hbar\omega$, $\delta \mu$, and $\delta T$, we find the explicit formulas for the $L_{ij}$ given in App.\ \ref{appcoeff}. One can also check that these expressions for the response coefficients satisfy the generalized Onsager relations Eq.\ (\ref{BL}), as they should.

\section{Efficiency and figure of merit of quantum machines}\label{seceff}
\subsection{Motors and generators}

Consider a situation with applied {\em ac} driving forces and a {\em dc} bias $\delta \mu$, but uniform temperature $T$. The device in Fig.\ \ref{setup}(a) can operate as a quantum motor or generator. When the {\em ac} potentials pump particles into the reservoir with lower chemical potential, the gain in electrical energy can be used to perform work on the source of the {\em ac} potentials. This occurs for $L_{31} \delta\mu/T<0$ and corresponds to a motor as the work performed on the {\em ac} potentials can be further transformed, say, into mechanical work. \cite{bus} When reversing the sign of $\delta\mu$ and thus $L_{31} \delta\mu/T>0$, the {\em ac} potentials pump particles into the reservoir with higher chemical potential and we have a generator. 

Using $X_2=0$, the rate of entropy production becomes 
\begin{equation} \label{sll}
  \overline{\dot{ S}}=  L_{11} X_1^2+L_{33} X_3^2+\left(L_{13}+ L_{31}\right)X_1 X_3.
\end{equation}
Interestingly, the last term on the right-hand side vanishes due to the Onsager symmetry $L_{13}=-L_{31}$, and the coefficients $L_{13}$ and $L_{31}$ do not affect the entropy production. As a consequence, the second law of thermodynamics imposes $L_{11}>0$ and $L_{33}>0$. This is in contrast to conventional thermoelectric, where the off-diagonal  response coefficients are symmetric, $L_{12}=L_{21}$, and do contribute to entropy production. In the latter case, the second law imposes ${\rm det} L = L_{11}L_{22}-L_{12}^2 > 0$ in addition. 
 
We are now ready to characterize the performance of adiabatically-driven quantum motors or generators in terms of efficiencies and figures of merit. The efficiency $\eta^{\rm mot}$ of a motor is measured by the ratio of the work per unit time $-\overline{\dot W}$ performed on the {\em ac} potentials and the  power $\overline{\dot{N}}_ R \delta \mu/e$ injected by the voltage source. Similarly, the efficiency of the generator $\eta^{\rm gen} $ is given by the inverse of this ratio, so that  
\begin{equation}\label{effmotapp}
\eta^{\rm mot} = \frac{1}{\eta^{\rm gen}}=\frac{-\overline{\dot{W}}}{\overline{\dot{N}}_ R \delta \mu/e}.
\end{equation}
Note that we have defined  $\mu_L=\mu_R+\delta\mu$ (as well as $T_L=T_R=T$), cf.\ Fig.\ \ref{setup}.

We first show that the second law of thermodynamics implies an upper limit for these efficiencies. Using Eqs.\ (\ref{sac}), (\ref{heat}), and (\ref{cons}), we find 
\begin{equation}\label{defje2}
  \overline{\dot{W}}=T\overline{{\dot{S}}}-\frac{\delta\mu}{e} \overline{\dot{N}}_R.
\end{equation}
Substituting this into Eq.\ (\ref{effmotapp}), we obtain
\begin{equation} \label{effdef1}
\eta^{\rm mot} = \frac{1}{\eta^{\rm gen}}=  1-\frac{T\overline{\dot{S}}}{\overline{\dot{N}}_ R \delta \mu/e}.
\end{equation}
Now, the second law of thermodynamics demands $\overline{\dot{S}}>0$. Moreover, the current flows with the potential drop $\delta\mu$ in the motor, $\overline{\dot{N}}_ R \delta \mu>0$, but against the potential drop for a generator, $\overline{\dot{N}}_ R \delta \mu<0$. Consequently, we find that both $\eta^{\rm mot}$ and $\eta^{\rm gen}$ are upper bounded by unity.

The efficiency in Eq.\ (\ref{effmotapp}) can also be written as
\begin{equation}\label{effmotapp2}
\eta^{\rm mot} = \frac{1}{\eta^{\rm gen}}=-\frac{X_3 J_3}{X_1 J_1}
\end{equation}
and the currents expressed through their linear-response expressions (\ref{ont}). Still assuming time-reversal symmetry, so that $L_{13}=-L_{31}$, we can then maximize the efficiency as a function of $X_1$ at fixed $X_3$. One finds that the efficiency is maximized for 
\begin{equation}\label{x1max}
X_{1}=\frac{L_{11}L_{33}\pm\sqrt{L_{11}L_{33}\text{det} {L}}}{L_{11}L_{13}}X_3.
\end{equation}
with $+(-)$  for  motors (generators). Alternatively, we can fix $X_1$ is fixed and maximize the efficiency as a function of $X_3$. In this case, one finds 
\begin{equation}\label{x2max}
X_3=\frac{-L_{11}L_{33}\pm\sqrt{L_{11}L_{33}\text{det} {L}}}{L_{33}L_{13}}X_1,
\end{equation}
again with $+(-)$  for  motors (generators). Substituting Eqs.\ (\ref{x1max}) and (\ref{x2max}) into Eq.\ (\ref{effmotapp2}), we find for the maximal efficiency
\begin{equation}
  \label{effmax1}
  \eta^{\rm max} = \frac{\sqrt{1+ \zeta}-1}{\sqrt{1+ \zeta}+1}
\end{equation}
and identify the figure of merit as
\begin{equation} 
\label{figmet}
 \zeta = \frac{ - L_{13}  L_{31}}{L_{11} L_{33}}.
\end{equation}
Note that $\eta^{\rm max}$ and $\zeta$ are valid for both motors and generators.

Equations (\ref{effmax1}) and (\ref{figmet}) should be contrasted with conventional thermoelectrics, \cite{revcasati,ben,seif,ora,zt} where the optimal efficiency satisfies an analogous expression. In conventional thermoelectrics, the efficiency of converting heat into electrical energy is limited by the Carnot efficiency $\eta_C$. However, the maximal efficiency which can be reached given a set of linear response coefficients $L_{ij}$ is lower than the Carnot efficiency by a factor involving the figure of merit $ZT = L_{12}^2/{\rm det} L$, see Eq.\ (\ref{maxeff}). In contrast, Eq.\ (\ref{effmax1}) describes the efficiency of converting electrical energy into other (e.g., mechanical) forms of energy. This process is not fundamentally limited and hence Eq.\ (\ref{effmax1}) does not contain an analog of the Carnot efficiency. However, it still contains an analog of the factor involving the $ZT$ value, which contains an appropriate figure of merit $\zeta$. Thus, the motor efficiency $\eta^{\rm mot}$ is bounded by unity, and reaches this limit when $\zeta\to\infty$, \textit{i.e.}, when one of the dissipative coefficients $L_{11}$ or $L_{33}$ approaches zero. The different form of the figure of merit, i.e., the absence of the coefficients  $L_{13}$ and $L_{31}$ from the denominator, reflects the fact that unlike  $L_{12}$ and $L_{21}$, these coefficients do not affect entropy production.

\subsection{Heat engine and heat pump}

Analogous results are obtained when the device is driven by a temperature gradient $\delta T$ at constant chemical potential ($X_1=0$), see\ Fig.\ \ref{setup}(b). When the device operates as a heat engine, \textit{i.e.}, for $L_{32}\delta T/T^2 <0$, heat flows to the cold reservoir and the system performs work on the {\em ac} potentials. Conversely, the device operates as a heat pump when $L_{32}\delta T/T^2 >0$, where heat is pumped to the hot reservoir by the {\em ac} potentials. As a result of the Onsager symmetry, we have $L_{23}=-L_{32}$ for time-reversal symmetric systems, and we again find that the second law imposes $L_{22}>0$ and $L_{33}>0$. 

An appropriate measure of the efficiency of a heat engine $\eta^{\rm he}$ is the ratio of the work per unit time performed by the electrons on the {\em ac} forces, $-\overline{\dot W}$, and the heat leaving the hot reservoir $-\dot{Q}_L$. (We assume that the left reservoir with temperature $T_R=T_L-\delta T$ is the hot reservoir. The efficiency $\eta^{\rm hp}$ of a heat pump is characterized by the inverse ratio. Thus, we have 
\begin{equation}\label{effheapp}
\eta^{\rm he} = \frac{1}{\eta^{\rm hp}}= \frac{\overline{\dot{W}}}{\overline{\dot{Q}}_L}
\end{equation}
for the efficiencies of heat engine and heat pump.

We first show that the second law implies that these efficiencies are bounded by the corresponding Carnot efficiencies. Using Eqs.\ (\ref{sac}), (\ref{heat}), and (\ref{cons}), we find 
\begin{equation}\label{defQL2}
  \overline{\dot{W}} =  T_R\overline{\dot{S}} + \frac{T_L-T_R}{T_L}\overline{\dot{Q}}_L.
\end{equation}
Inserting this into the definition (\ref{effheapp}) of the efficiencies, we obtain
\begin{equation}
  \eta^{\rm he} = \frac{1}{\eta^{\rm hp}} = \frac{T_L-T_R}{T_L} + \frac{T_R\overline{\dot{S}}}{\overline{\dot{Q}}_L}.
\end{equation}
For heat engines, heat flows from the hot to the cold reservoir so that $\overline{\dot{Q}}_L <0$, while for heat pumps, heat flows in the opposite direction, $\overline{\dot{Q}}_L>0$. Thus, we find that the second law $\overline{\dot{S}}>0$ implies that the efficiencies are smaller than the familiar Carnot efficiencies, i.e., $\eta_C = {(T_L-T_R)}/{T_L}$ for the heat engine and  $\eta_C = {T_L}/{(T_L-T_R)}$ for the heat pump. 

The efficiencies for heat engine and heat pumps can alternatively be expressed as
\begin{equation}\label{effdef2a}
\eta^{\rm he} = \frac{1}{\eta^{\rm hp}}=-\frac{X_3 J_3}{X_2J_2},
\end{equation}
where the fluxes can be expressed through their linear-response expressions (\ref{ont}). Maximizing the efficiency as for motors and generators, we again find Eqs. \ (\ref{x1max}, \ref{x2max}), but with the affinity $X_2$ taking the place of $X_1$. This leads to a maximal efficiency of
\begin{equation}\label{effmax2}
   \eta^{\rm max} = \eta_c \frac{\sqrt{1+ \tilde{\zeta}}-1}{\sqrt{1+ \tilde{\zeta}}+1}
\end{equation}
with the figure of merit
\begin{equation} \label{figmet2}
   \tilde{\zeta}= \frac{ -  L_{23} L_{32}}{L_{22} L_{33}}.
\end{equation}
Equation (\ref{effmax2}) holds for both heat engines and heat pumps, when the appropriate Carnot efficiency $\eta_c$ is used. 

\begin{figure}[t]\begin{center}
  \includegraphics[scale=0.4]{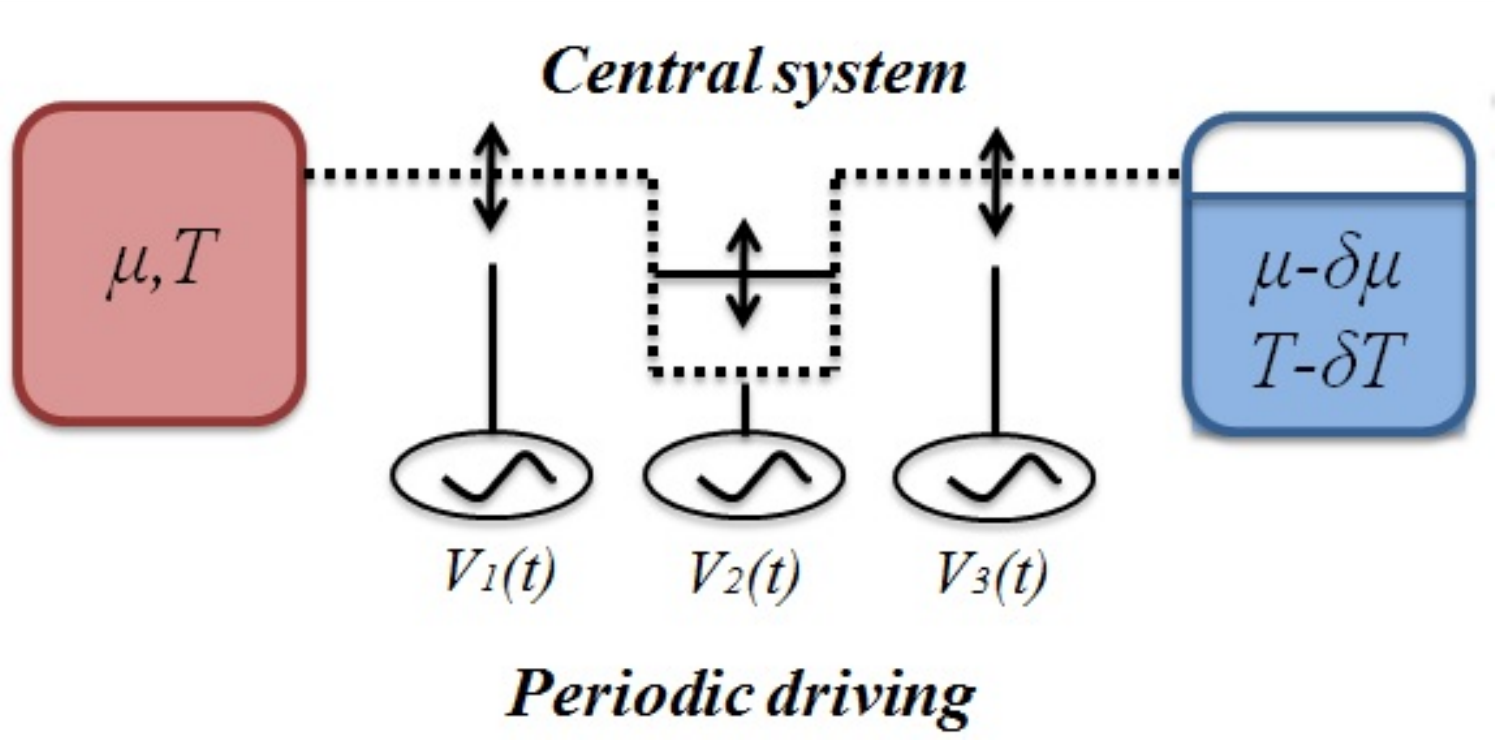}\\
  \caption{ Sketch of the device. A single-level quantum dot  ($m=2$) is defined by two tunnel barriers ($m=1,3$). They are driven by periodic gate potentials 
   $V_m(t)=V_j^0 \cos(\omega t + \delta_m)$, with $V_1^0=V_3^0=4$, $V_2^{0}=23$, $\delta_1=0$, $\delta_2=\pi/2$, and $\delta_3=\pi$. The tunneling amplitudes between barriers and dot are $w=1$ and  $w_L=w_R=0.7$ between barriers and reservoirs. The barriers and the dot are modeled by a discrete chain of $N=3$ sites with local energies $\varepsilon_1=\varepsilon_3=3.3$ and $\varepsilon_2=-1$ respectively.   The reservoirs have $\mu_L=\mu, \; \mu_R=\mu-\delta\mu$ and temperature $T$. }\label{plota}
\end{center}\end{figure}

\section{Example}
\label{secex}

To illustrate these concepts, we consider a quantum dot with a single level coupled to two reservoirs  with chemical potentials $\mu_{\alpha}$ and temperatures $T_{\alpha}$, $\alpha=L,R$ as sketched in Fig.\ \ref{plota}. We assume that the dot level and the barriers can be modulated periodically in time by {\em ac} gate potentials. This model can describe a single-electron source, similar to the GHz pump realized experimentally in Ref.\ \onlinecite{highfreq}. For noninteracting electrons, the model is described by the Hamiltonian
\begin{equation}
\hat{{\cal H}}(t)=\hat{{\cal H}}_c(t)+\hat{{\cal H}}_{res}+\hat{{\cal H}}_T.
\end{equation}
The first term  describes the central conductor which is modeled as a discrete chain of $N$ sites  with local energies $\ve_m$, nearest-neighbor hopping $w$, and an {\em ac} potential applied to each site, 
\beq \label{hamph}
\hat{{\cal H}}_c(t)\!=\!\sum_{m=1}^N\!\left[\!\left(\!\ve_m + V_m(t) \right) d^{\dagger}_m d_m+\sum\limits_{ m=1}^{N-1} wd^{\dagger}_{m} d_{m+1}\!\right]\!+\!\mbox{h.c.}
\edq
Specifically, we consider a setup with $N=3$ sites, modeling the tunneling barriers ($m=1,3$) and the quantum dot ($m=2$). The site energies $\varepsilon_m$ ($m=1,2,3$) are modulated by three time-dependent gate voltages of the form $V_m(t)=V_m^0 \cos(\omega t + \delta_m)$. The reservoirs are represented by free-electron Hamiltonians for free electrons, 
\beq
\hat{{\cal H}}_{res}=\sum_{{\alpha}=L,R,{k_\alpha}} E_{k_\alpha}c^{\dagger}_{k_\alpha}c_{k_\alpha},
\edq
and tunneling between reservoirs and central system is described by
\beq \label{hamcont}
\hat{{\cal H}}_T=-\sum_{\alpha,k_\alpha,n} [w_{\alpha} d^{\dagger}_{n_{\alpha}} c_{k_\alpha} + h.c],
\edq
where $n_{\alpha}$ denotes the site of the central conductor which is in contact with the reservoir $\alpha$.

The mean charge current ${\overline{\dot{N}}_{\alpha}}$ and heat current $\overline{\dot{Q}}_{\alpha}$ entering the reservoir $\alpha$, as well as the mean power $\overline{\dot{W}}$ developed by the {\em ac} forces are calculated within a Floquet Green function formalism following Ref.\ \onlinecite{arr07}, as reviewed in App.\ \ref{green}. To derive the response coefficients $L_{ij}$, we expand the currents $J_1= \overline{\dot{N}}_R$,  $J_2= \overline{\dot{Q}}_R$, and $J_3= \overline{\dot{W}}/(\hbar \omega)$ to linear order in $\hbar \omega$, see App.\ \ref{applinres}. Explicit expressions for the coefficients $L_{ij}$ -- in terms of Green functions \cite{arr07,arr06,ludo2} or scattering matrices \cite{Moskalets04} and valid for noninteracting systems -- can be found in App.\ \ref{appcoeff}. These coefficients can also be calculated using an alternative procedure which does not  rely on the Floquet decomposition, see  Refs.\ \onlinecite{niels1,niels2}. However, we prefer to use the Floquet approach because this representation stresses that $\hbar \omega$ appears in the Fermi functions which enter the integrals for the currents on the same footing as the chemical potential $\mu$. This provides an alternative argument for identifying $\hbar \omega/T$ as an affinity.

\begin{figure}[t]\begin{center}
  \includegraphics[scale=0.4]{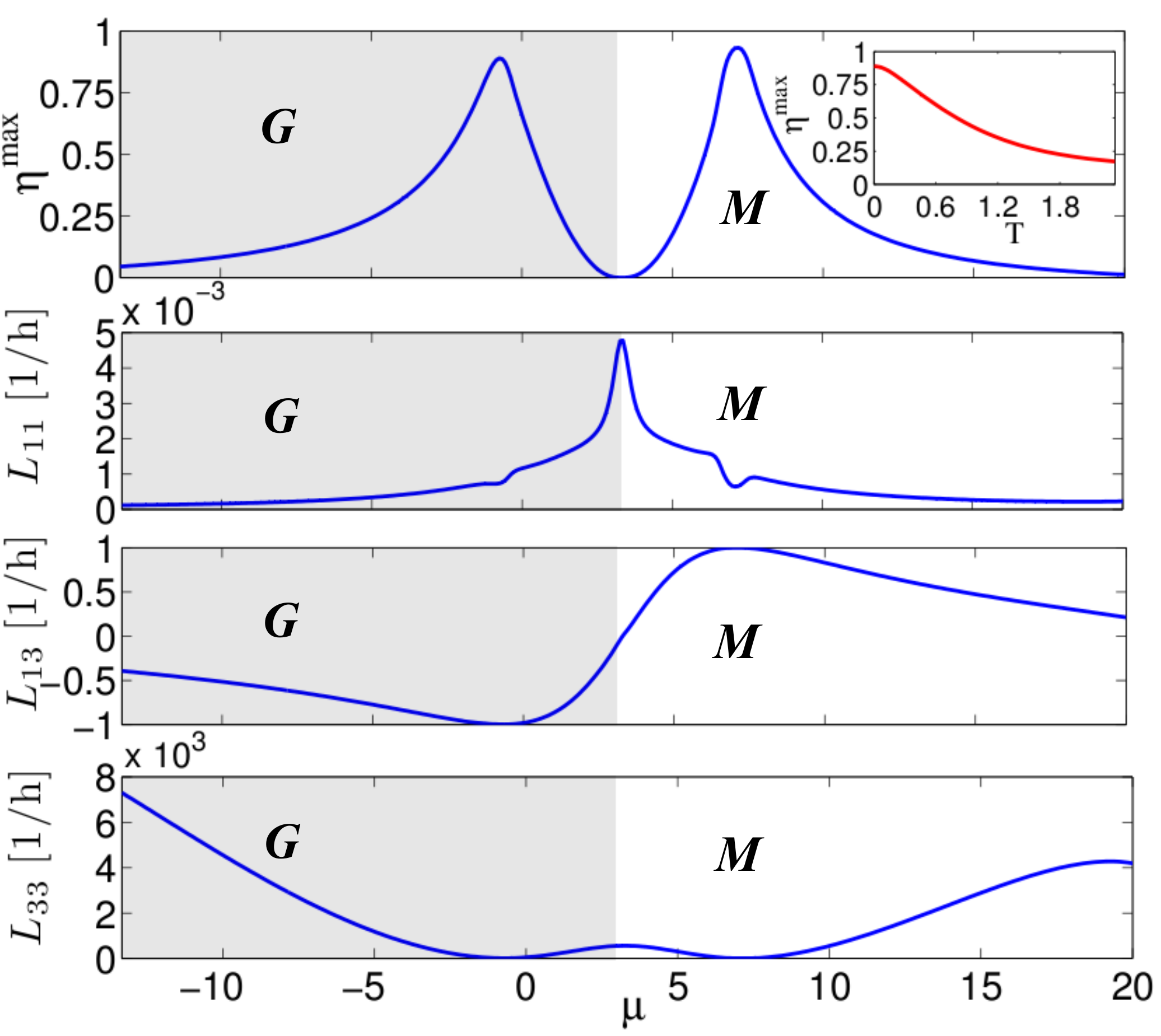}\\
  \caption{ Maximum efficiency $\eta^{\rm max}$ and 
   transport coefficients at $T=0$  for the  motor ($M$) or generator ($G$) modes. Inset: $\eta^{\rm max}$ for generator/motor with $\mu=-0.7$ $(\mu=7.2)$.  }\label{plot}
\end{center}\end{figure}

For illustration, we consider an applied bias $\delta \mu$ at $T=0$, \textit{i.e.}, the motor/generator regime, and calculate the coefficients listed in App.\ \ref{appcoeff} for this case. In Fig.\ \ref{plot}, we plot the transport coefficients  and the maximum efficiency $\eta^{\rm max}$ as functions of the chemical potential $\mu$ of the left reservoir. Large values of the figure of merit require a large charge pumping coefficient $L_{13}$ along with a small value of $L_{33} L_{11}$, \textit{i.e.}, low friction or conductance. In the absence of  driving at the central dot [$V_2(t)=0$], the conductance peaks near $L_{11}=1$ when $\mu$ is in resonance with the dot level. Driving the dot level with a phase lag relative to the barrier oscillations ($\delta_2-\delta_m\neq 0$ for $m=1,3$) favors charge pumping and decreases the conductance by dynamically tuning the dot off resonance. In this way, high efficiencies can be achieved despite large values of $L_{33}$. 

As the chemical potential passes the dot level, the pumping coefficient changes sign, and the system switches from motor mode [$L_{31}\delta\mu/T<0$; see region $M$ in the Fig.\ \ref{plot}] to generator mode [$L_{31}\delta\mu/T>0$, see region $G$ in the Fig.\ \ref{plot}]. The efficiency becomes minimal when the chemical potential is resonant with the dot level, where the conductance is maximal and pumping vanishes by particle-hole symmetry. 
 
The device can also operate as a heat engine or pump when imposing a temperature gradient. As this requires finite $T$, quantum effects are less pronounced and efficiencies are lower than those shown in Fig.\ \ref{plot}. However, we find that for appropriate parameters these may still be as high as $\approx 0.4 \eta_c$.

\section{Summary}\label{seccon}

Motivated in part by Jarzynski's equality \cite{jarz} and Crook's theorem, \cite{crook} there has been much interest in quantum thermodynamics, including fluctuation relations, work fluctuations, and the thermodynamic description of strongly coupled systems. \cite{kur,rev1,rev2} Here, we provided a generalized thermoelectric framework to analyze the thermodynamics of {\em ac}-driven nanoscale systems which explicitly accounts for the effects of quantum pumping and the related nonconservative forces. We identified the additional flux and affinity through which these forces enter the theory and defined generalized Onsager relations for the associated response coefficients. This framework allowed us to define appropriate efficiencies and figures of merit which describe quantum motors, generators, heat engines, and heat pumps. We illustrated our results for a simple quantum-pump device. 

\section{Acknowledgments}
We acknowledge useful discussions with A.\ Bruch, R.\ Bustos-Marun, S.\ Kusminskiy, and A.\ Nitzan. This work was supported by the Alexander-von-Humboldt Stiftung (L.A. and F.v.O.), CONICET (M.F.L. and F.B.), MINCyT and UBACyT (L.A.), as well as the Deutsche Forschungsgemeinschaft (F.v.O.).

\appendix

\section{Response coefficients and microreversability}\label{apon}
The matrix elements entering ${\bf \Lambda}^{cf}_t$ read
 \ba\label{cft}
 \Lambda^{cf}_j&=&-i\int_{-\infty}^{\infty} dt^{\prime} (t-t^{\prime}) \theta(t-t^{\prime}) \langle\left[\hat{J}^c(t), \hat{F}_j(t^{\prime})\right]\rangle_t \nonumber \\ 
 &\equiv & \int_{-\infty}^{\infty} d\tau \tau
 \chi^{J,F_j}_t(\tau),
 \ea
 where we have defined the retarded susceptibility $\chi^{J,F_j}_t(\tau) = - i \theta(\tau)\langle \left[\hat{J}^c(\tau), \hat{F}_j(0)  \right]\rangle_t$. In the latter step we have stressed that for evolutions with the operator $U(\tau)= e^{-i \tau \hat{H}_t}$, being $\hat{H}_t= \hat{H}\left({\bf V}(t)\right)$ the frozen Hamiltonian, the actual time argument of the integrand of
 (\ref{cft}) is $\tau=t-t^{\prime}$. Representing the susceptibility in terms of the Fourier transform  we can also write the previous expression as
\ba \label{cft1}
 \Lambda^{cf}_j&=&\mbox{Re}\left[\int_{-\infty}^{+\infty}  d\tau \!\!\int_{-\infty}^{+\infty}   \frac{d\omega}{2 \pi  i} e^{-i \omega \tau} \partial_{\omega} \chi^{J,F_j}_t(\omega)\right]= \nonumber \\ 
 &=& \mbox{Re}\left[ -i \int_{-\infty}^{+\infty}  d\omega \partial_{\omega} \chi^{J,F_j}_t(\omega) \delta(\omega) \right]=\nonumber \\ 
 &=&   \lim_{\omega \rightarrow 0} \frac{\mbox{Im}\left[ \chi^{J,F_j}_t(\omega) \right]}{\omega},
 \ea
 where we have used that  the spectral function $\mbox{Im}\left[ \chi^{J,F_j}_t(\omega) \right]$ is odd in $\omega$, hence 
 $\mbox{Im}\left[ \chi^{J,F_j}_t(0) \right] =0$. \cite{bruus}
 
 Analogously, the matrix elements of  $\hat{\Lambda}^{ff}_t$ can be written as
   \begin{eqnarray}
 \Lambda^{ff}_{ij} &=&  -i \int_{-\infty}^{\infty} dt^{\prime} (t-t^{\prime}) \theta(t-t^{\prime}) \langle \left[\hat{F}_i(t), \hat{F}_j(t^{\prime})  \right]\rangle_t= \nonumber \\
&=&  \mbox{Re}\left[ -i \int_{-\infty}^{+\infty}  d\omega \partial_{\omega} \chi^{F_i,F_j}_t(\omega) \delta(\omega) \right]= \nonumber \\
 & = & \lim_{\omega \rightarrow 0} \frac{\mbox{Im} \left[ \chi^{F_i,F_j}_t (\omega)\right]}{\omega} ,
 \end{eqnarray}
 where $\chi^{F_i,F_j}_t (\omega)$ is the Fourier transform of  $\chi^{F_i,F_j}_t(\tau) = - i \theta(\tau)\langle \left[\hat{F}_i(\tau), \hat{F}_j(0)  \right]\rangle_t$. 
 
 The calculation of the conductivity follows the usual procedure of the Kubo formula presented in text books. \cite{bruus} We start by considering an extra perturbation due to the coupling to an electric field $E(t)=\partial_t A(t)$. In the Fourier domain the extra perturbation is
 $
 {\cal H}^{\prime}(\omega)=  {\cal J} \cdot E(\omega)/ (i \omega)$,
  which leads to the definition of the {\em dc} conductance
  \begin{equation}
  \Lambda^{cc}=  \lim_{\omega \rightarrow 0}   \frac{ \mbox{Im}\left[ \chi^{J,J}_t (\omega) \right]}{\omega},
  \end{equation}
  where  $\chi^{J,J}_t (\omega)$ is the Fourier transform of $\chi^{J,J}_t(\tau) = - i \theta(\tau)\langle \left[\hat{J}(\tau), \hat{J}(0)  \right]\rangle_t$.  
  
  Similarly, evaluating the forces in linear response with respect to $\delta \mu$ leads to
  \begin{equation}
   \Lambda^{fc}_j =   \lim_{\omega \rightarrow 0}   \frac{ \mbox{Im}\left[  \chi^{F_j,J}_t (\omega)\right]}{\omega},
  \end{equation}
  where  $\chi^{F_j,J}_t (\omega)$ is the Fourier transform of $\chi^{F_j,J}_t(\tau) = - i \theta(\tau)\langle \left[\hat{F}_j(\tau), \hat{J}(0)  \right]\rangle_t$.  
  
 The above definitions indicate that the susceptibilities $\chi^{O_i,O_j}_t$, with $\hat{O}_i$ a generic operator, satisfy microreversibility with respect to $\tau$. It can be directly verified that
 \ba 
\chi^{O_i,O_j}_t(-\tau) &=&- i  \theta(-\tau) \langle \left[\hat{O}_i(-\tau), \hat{O}_j(0)  \right]\rangle_t= \nonumber\\
&=& i  \theta(-\tau) \langle \left[\hat{O}_j(\tau), \hat{O}_i(0)  \right]\rangle_t=  \\\nonumber
&=&{-}i \theta(-\tau) \int_{-\infty}^{+\infty} \frac{d \omega}{\pi} \mbox{Im}[\chi^{O_j,O_i}_t (\omega)] e^{-i \omega \tau}.
\ea   
   Hence, under a transformation $\tau \rightarrow -\tau$ the  coefficient $\Lambda_{ij}$ transforms to
   \ba \label{tranlan}
\Lambda^{O_i,O_j}_{ij}\!\!\!\!& =&\mbox{Re}\!\!\left[\!i\!\! \int_{-\infty}^{+\infty}\!\!\frac{d \omega}{\pi}\!\mbox{Im}[\!\chi^{O_j,O_i}_t (\omega)]\!\!\int_{-\infty}^{+\infty}\!\!\!\!d\tau \tau \theta(-\tau)e^{-i \omega \tau}\!\right]\! \nonumber \\
&=&\lim_{\omega \rightarrow 0}\frac{\mbox{Im}\left[ \chi^{O_j,O_i}_t (\omega)\right]}{\omega}= \Lambda^{O_j,O_i}_{ji}.
  \ea
  In the last step we have used 
% \begin{equation}
$\int_{-\infty}^0 d\tau \tau e^{-i \omega \tau}= \frac{1}{\omega^2}+ i \pi \delta^{\prime}(\omega)$.
%\end{equation}

In the presence of a magnetic field $B$, a time-reversal transformation implies changing $B \rightarrow -B$ in the Hamiltonian ${\cal H}_t$ defining the frozen density matrix $\hat{\rho}_t$ used to evaluate the expectation values. This property leads to the following Onsager relations for the usual susceptibilities in the presence of $B$,
$\chi^{O_i,O_j}_t(B, \omega)=  s_i s_j  \chi^{O_j,O_i}_t(-B, \omega)$, where the signs $s_i, s_j = \pm$ depend on the parity of the operators $\hat{O}_i, \; \hat{O}_j$ under
a time-reversal transformation. 

\section{Green function formalism}\label{green}
In Ref. \onlinecite{arr07} it was shown that the averaged charge ${\overline{\dot{N}}_{\alpha}}$ and heat  $\overline{\dot{Q}}_{\alpha}$ currents entering the reservoir $\alpha$,  
as well as the mean power $\overline{\dot{W}}$ developed by the ac forces
can be written in terms of the retarded  Green function of the central structure connected to the reservoirs expanded in the Floquet-Fourier transform as
\beq \label{gr}
\hat{G^R}(t,t^{\prime})\!=\!\sum_{n=-\infty}^{\infty}e^{-i n \omega t}\!\int_{-\infty}^{\infty}\!\!\frac{d E}{2 \pi}  e^{-i \frac{E}{\hbar}(t-t^{\prime})}\hat{\cal G}(n,E).
\edq
This function is calculated from the Hamiltonian $\hat{\cal H}(t)$ by solving the Dyson equation (see Refs. \onlinecite{arr06,arr07}). 

The resulting expression for the charge current is
\ba
\label{charge}
\overline{\dot{N}}_{\alpha}\!=\!\frac{e}{h}\!\int\!\!\!{dE}\sum_{n,\beta}\!\left[f_\beta(E)\!\!-\!\!f_\alpha(E+n\hbar\omega)\!\right]\!{\cal T}_{\alpha \beta}(n,E),
\ea
with
\begin{equation}
{\cal T}_{\alpha \beta}(n,E)=\vert \hat{\cal G}_{\alpha \beta}(n,E)\vert ^2\hat{\Gamma}_\beta\hat{\Gamma}_\alpha. 
\end{equation}
The heat current reads
\ba
\label{energy}
\overline{\dot{Q}}_\alpha=\overline{\dot{E}}_\alpha-\mu_\alpha \frac{\overline{\dot{N}}_\alpha}{e},
\ea
with
\ba
\overline{\dot{E}}_\alpha\!=\!\!\!\int \!\!\!{dE}\sum_{n} \frac{E}{h}  [f_\beta(E)\!\!-\!\!f_\alpha(E+n\hbar\omega)]\!{\cal T}_{\alpha \beta}(n,E).
%\vert\hat{\cal G}_{\alpha\beta}(n,E)\vert^2\hat{\Gamma}_\beta\hat{\Gamma} _\alpha.
\ea
Similarly,  the work performed by the {\em ac}  potentials can be written as
\ba\label{work}
\overline{\dot{W}}&=&-\frac{1}{h}\!\!\sum_{\alpha,l,n}\int_{-\infty}^{+\infty}\!\!\!{dE}  n  \hbar\omega f_{\alpha}(E) \nonumber       \\
& &\times \text{Im}\!\left\{\text{Tr}\!\left[\hat{V}(n)\hat{\cal G}(n+l,E)\hat{\Gamma}_{\alpha} \hat{\cal G}^{\dagger}(l,E)\!\right]\!\right\}, \\\nonumber
\ea
where $\hat{V}(n)$ are the Fourier components of $\hat{V}(t)=\sum_n \hat{V}(n)e^{in\omega t}$, being $\hat{V}(t)$ a matrix with diagonal elements $V_m(t)$.
In the above expressions we introduced the hybridization matrix $\hat{\Gamma}_{\alpha}$ which has a single element at the contact with the reservoir equal to $|w_{\alpha}|^2 2 \pi \sum_{k_{\alpha}} \delta(E-E_{k_{\alpha}}) $. For practical uses it can be considered in the wide band limit, thus, independent of $E$.  The Fermi-Dirac distribution $f_\alpha(E)=[1+e^{(E-\mu_\alpha)/T_\alpha}]^{-1}$ characterizes the thermal occupation of the electrons in the reservoirs (from now on we set the Boltzmann constant $k_B=1$).

The other possible approach is the Floquet scattering matrix formalism used in Ref. \onlinecite{Moskalets04}. The elements $s_{ij}(E_m, E_n)$ of the Floquet scattering matrix $\hat{s}(E)$, with $E_n=E+n\hbar\omega$,
are the amplitudes for an electron to scatter from lead $j$ to lead $i$ after acquiring $m-n$ Floquet quanta $\hbar \omega$. 
 The general relation between the Floquet scattering matrix elements and the Fourier coefficients for the Green's function is the generalized Fisher-Lee formula  \cite{arr06} 
\begin{equation}\label{relation}
s_{ij}(E_m, E_n)=\delta_{ij}\delta_{mn}-i\sqrt{\Gamma_i\Gamma_j}{\cal G}_{ij}(m-n, E_n).
\end{equation}

\section{Linear response}\label{applinres}
In order to calculate the currents $J_l, \; l=1,2,3$ up to linear order in $\hbar \omega$,  $\delta \mu$ and $\delta T$ we perform the following expansion
of the Fermi function entering the integrands of Eqs. (\ref{charge}), (\ref{energy}) and Eq.\ (\ref{work})
\ba\label{Fermiexp}
f_{\alpha}(E&+&n\hbar\omega)\sim f_{\alpha}(E)+n\hbar\omega\partial_E f_{\alpha}(E)  \\\nonumber
& &-\frac{\partial f(E)}{\partial E}(\mu_{\alpha}-\mu) -\frac{\partial f(E)}{\partial E}\frac{(E-\mu)}{T}(T_{\alpha}-T).
\ea

We also evaluate ${\cal G}(n,E)$ up to linear order in $\omega$ by expanding the Dyson equation  in powers of $\omega$ (see \onlinecite{ludo1,ludo2}). Up to the first order in $\omega$ it reads
\ba\label{gfroz}
\hat{G}(t,E) \sim \hat{G^f}(t,E)+  i\hat{G^{(1)}}(t,E), \ea
with $\hat{G}(t,E)= \sum_{n=-\infty}^{\infty} \hat{\cal G}(n,E) e^{-i n \omega t} $. The first term is the frozen Green function
\beq
\hat{G^f}(t,E)=\left[\hat{1}\; E -\hat{\cal H}^t_c  + i \frac{\hat{\Gamma}}{2} \right]^{-1},
\edq
corresponding to the frozen Hamiltonian at time $t$,  $\hat{\cal H}^t_c = \hat{\cal H}_c(t)$ ($ \hat{\Gamma}$ collects the hybridization functions of the reservoirs). The next term is  first order in $\omega$. It reads
\begin{equation}
\hat{G^{(1)}}(t,E) =  \frac{\hbar}{2}\partial_E \partial_t\hat{G^f}(t,E)+\hat{\mathcal{A}}(t,E), 
\end{equation}
where 
\beq\label{a}
\hat{\mathcal{A}}\!\!=\!\!\frac{\hbar}{2}\!\!\left(\!\!\partial_E \hat{G^f}(t,E)\frac{d\hat{V}}{dt}\hat{G^f}(t,E)\!-\!\hat{G^f}(t,E)\frac{d\hat{V}}{dt}\partial_E \hat{G^f}(t,E)\!\!\right)\!.
\edq
The expansion of  the  Floquet scattering matrix
  up to the first order in the driving frequency $\omega$ cast
\ba\label{frozen}
 s_{ij}(E, E_n)&=&\frac{1}{\tau}\int_0^{\tau}dt \mbox{ }e^{-in\omega t}[s_{ij}(t,E)+  \\\nonumber
&+&\frac{n\hbar\omega}{2}\partial_E s_{ij}(t,E)+\hbar\omega A_{ij}(t,E)].
\ea
Here $s_{ij}(t,E)$ is the frozen scattering matrix. 
 The matrix elements $A_{ij}(t,E)$ define a first order  correction to the adiabatic scattering matrix.  The frozen scattering matrix $s_{ij}(t,E)$ as well as $A_{ij}(t,E)$  do not change significantly on the
energy scale $\hbar\omega$  and $T$ and depend on the specific realization of the scatterer. Anyway, it can be shown that, due to the unitarity of the Floquet scattering matrix and of the frozen scattering matrix they satisfy \cite{Moskalets04}
\begin{equation}\label{amatrix}
\hbar\omega[\hat{s}^{\dagger}\hat{A}+\hat{A}^{\dagger}\hat{s}]=\frac{i\hbar}{2}\left(\frac{\partial \hat{s}^{\dagger}}{\partial t}\frac{\partial \hat{s}}{\partial E}-\frac{\partial \hat{s}^{\dagger}}{\partial E}\frac{\partial \hat{s}}{\partial t}\right).
\end{equation}
Equation (\ref{relation}) defines an explicit relation between $\hat{\cal A}$ and $\hat{A}$. \cite{arr06}

\section{Transport coefficients}\label{appcoeff}

Substituting  the expansions for the Fermi function, Eq.\  (\ref{Fermiexp}), and for the Green function, Eq.\ (\ref{gfroz}), into Eqs. (\ref{charge}, \ref{energy}) and Eq.\ (\ref{work}) and 
collecting terms up to first order in the affinities $X_1=\frac{\delta\mu}{T}$, $X_2=\frac{\delta T}{T^2}$  and $X_3=\frac{\hbar\omega}{T}$ we obtain:
\begin{widetext}
\ba\label{excoeff}
L_{11} & = &-\frac{T}{h{\cal T}}\int^{\cal T}_0dt\int_{-\infty}^{+\infty} {dE}\frac{df}{dE}\vert\hat{G}^{f}_{RL}(t,E)\vert^{2}\hat{\Gamma}_L\hat{\Gamma}_R\nonumber\\
L_{12} & = &  L_{21}=-\frac{T}{h{\cal T}}\int^{\cal T}_0dt\int_{-\infty}^{+\infty} {dE}(E-\mu)\frac{df}{dE}\vert\hat{G}^{f}_{RL}(t,E)\vert^{2}\hat{\Gamma}_L\hat{\Gamma}_R\nonumber\\
L_{13} & = &-L_{31}=-\frac{T}{2\pi h}\int^{\cal T}_0dt\int_{-\infty}^{+\infty} {dE}\frac{df}{dE}\mbox{Im}\left\{\left[\hat{G}^{f}(t,E)\hat{\Gamma}\frac{{\partial}\hat{G}^{f\dagger}(t,E)}{\partial t}\hat{\Gamma}\right]_{RR}\right\}\nonumber\\
{L}_{22} & = &-\frac{T}{h{\cal T}}\int^{\cal T}_0dt\int_{-\infty}^{+\infty} {dE}\,(E-\mu)^2\,\frac{df}{dE}\vert\hat{G}^{f}_{RL}(t,E)\vert^{2}\hat{\Gamma}_L\hat{\Gamma}_R\nonumber\\
{L}_{23} & = & -L_{32}=-\frac{T}{2 \pi h}\int^{\cal T}_0dt\int_{-\infty}^{+\infty} {dE}\,(E-\mu)\,\frac{df}{dE}\mbox{Im}\left\{\left[\hat{G}^{f}(t,E)\hat{\Gamma}\frac{\partial\hat{G}^{f\dagger}(t,E)}{\partial t}\hat{\Gamma}\right]_{RR}\right\}\nonumber\\
L_{33} & = & -\frac{T {\cal T} }{8\pi^2h}\int^{\cal T}_0dt\int_{-\infty}^{+\infty} {dE}\frac{df}{dE}
\mbox{Re}\left\{\mbox{Tr}\left[\frac{\partial\hat{G}^{f}(t,E)}{\partial t}\hat{\Gamma}\frac{\partial\hat{G}^{f\dagger}(t,E)}{\partial t}\hat{\Gamma}\right]\right\}. 
%\hat{\mathcal{A}}(t,E)\hat{\Gamma}\partial_t\hat{G}^{f\dagger}(t,E)\hat{\Gamma}\right] \} \right .\nonumber\\
%& & \left.+\frac{\hbar}{2}\mbox{Re}\{\mbox{Tr}\left[\partial_t\hat{G}^{f}(t,E)\hat{\Gamma}\partial_t\hat{G}^{f\dagger}(t,E)\hat{\Gamma}\right]\} \right].
\ea
Within the scattering matrix formalism the coefficients 
read
\begin{eqnarray}\label{coeff}\nonumber
{L}_{11}&=&-\frac{T}{ h{\cal T}}\int_0^{{\cal T}} dt\int_{-\infty}^{+\infty} dE \frac{df}{dE}|\hat{s}_{RL}(t,E)|^2\\\nonumber
{L}_{12}&=&L_{21}=-\frac{T}{h{\cal T}}\int_0^{{\cal T}} dt\int_{-\infty}^{+\infty} dE (E-\mu)\,\frac{df}{dE}|\hat{s}_{RL}(t,E)|^2\\\nonumber
{L}_{13}&=&-L_{31}=-\frac{T}{2\pi h}\int_0^{{\cal T}} dt\int_{-\infty}^{+\infty} dE \frac{df}{dE}\mbox{Im}\left\{\left[\hat{s}(t,E)\frac{\partial \hat{s}^{\dagger}(t,E)}{\partial t} \right]_{RR}\right\}\\\nonumber
{L}_{22}&=&-\frac{T}{h\tau}\int_0^{{\cal T}} dt\int_{-\infty}^{+\infty} dE (E-\mu)^2\,\frac{df}{dE} |\hat{s}_{RL}(t,E)|^2\\\nonumber
{L}_{23}&=&-L_{32}=-\frac{T}{2\pi h}\int_0^{{\cal T}} dt\int_{-\infty}^{+\infty} dE (E-\mu) \,\frac{df}{dE}\mbox{Im}\left\{\left[\hat{s}(t,E)\frac{\partial \hat{s}^{\dagger}(t,E)}{\partial t} \right]_{RR}\right\}\\\nonumber
{L}_{33}&=&
%&=&-\frac{T}{2\pi h}\int_0^{{\cal T}} dt\int_{-\infty}^{+\infty} dE E \partial_E [f(E)]\sum_{ij}2\mbox{Im}\left\{A_{ij}(t,E)\frac{\partial s_{ij}^{\ast}(t,E)}{\partial t}\right\}\\\nonumber
-\frac{T {\cal T} }{8\pi^2 h }\int_0^{{\cal T}} dt\int_{-\infty}^{+\infty} dE  \frac{df}{dE}\mbox{Tr}\left[\frac{\partial \hat{s}(t,E)}{\partial t} \frac{\partial \hat{s}^{\dagger}(t,E)}{\partial t}\right].\\
\end{eqnarray}
\end{widetext}
The matrices $\hat{\cal A}$ in the Green-function language, and  $\hat{A}$ in the scattering matrix version, in principle seem to contribute to the coefficient $L_{33}$.
In particular, they appear in an integrand of the form
\beq
\sum_{ij}2 \mbox{Im}\left\{A_{ij}(t,E)\frac{\partial s_{ij}^{\ast}(t,E)}{\partial t}\right\}.
\edq
However, as shown in Ref. \onlinecite{niels1},  due to the 
 unitary condition of the frozen scattering matrix $\hat{s}\hat{s}^\dagger=1$ and the property (\ref{amatrix}) such term vanishes. In fact, it can be also written as

\ba
%\mbox{Im}\left\{A_{ij}(t,E)\frac{\partial s_{ij}^{\ast}(t,E)}{\partial t}\right\}& = & 
& & 2\mbox{Im}\left\{\mbox{Tr}\left[{\partial_t}{\hat{s}^\dagger}\hat{A}\right]\right\}=-i\mbox{Tr}\left[{\partial_t}{\hat{s}^\dagger}\hat{A}-\hat{A}^\dagger{\partial_t}{\hat{s}}\right]=
  \\\nonumber
& &
 =-i\mbox{Tr}\left[\left({\hat{s}^\dagger}\hat{A}+\hat{A}^\dagger{\hat{s}}\right)\partial_t{\hat{s}^\dagger}\hat{s}\right]=
  \\\nonumber
& &
 =\frac{1}{2\omega}\mbox{Tr}\left[\left(\partial_t{\hat{s}^\dagger}\partial_E{\hat{s}}-{\hat{s}^\dagger}\partial_E{\hat{s}}\partial_t{\hat{s}^\dagger}{\hat{s}}\right)\partial_t{\hat{s}^\dagger}\hat{s}\right]
 =0.
\ea\\


\begin{thebibliography}{99}
\bibitem{footnote} We assume for definiteness that all parameters $V_j(t)$ transform in the same way under time reversal.
\end{thebibliography}
\end{document}